\documentclass[lettersize,journal]{IEEEtran}
\usepackage{amsmath,amsfonts}
\usepackage{algorithmic}
\usepackage{array}
\usepackage[caption=false,font=normalsize,labelfont=sf,textfont=sf]{subfig}
\usepackage{textcomp}
\usepackage{stfloats}
\usepackage{url}
\usepackage{verbatim}
\usepackage{graphicx}
\usepackage{balance}
\usepackage{color}
\usepackage{gensymb}
\usepackage{amsmath}
\usepackage{graphicx}

\begin{document}
    \title{Illumination-Based Color Reconstruction for the Dynamic Vision Sensor}
    
    \author{ \IEEEauthorblockN{
            Khen~Cohen, 
            Omer~Hershko,
            Homer~Levy,
            David~Mendlovic,
            and~Dan~Raviv 
            }
            }
    
    \maketitle

    \begin{abstract}
        This work demonstrates a novel, state-of-the-art method to reconstruct colored images via the Dynamic Vision Sensor (DVS). The DVS is an image sensor that indicates only a binary change in brightness, with no information about the captured wavelength (color), or intensity level. We present a novel method to reconstruct a full spatial resolution colored image with the DVS and an active colored light source. We analyze the DVS response and present two reconstruction algorithms: Linear-based and Convolutional-Neural-Network-Based. In addition, we demonstrate our algorithm's robustness to changes in environmental conditions such as illumination and distance. Finally, comparing with previous works, we show how we reach the state-of-the-art results.
    \end{abstract}

    \IEEEpeerreviewmaketitle

    \section{Introduction}
        %
        %
        %
        %
        \IEEEPARstart{T}{he} majority of the image sensors used nowadays is composed of CCD and CMOS image sensors \cite{CCD-CMOS-Market}. When using these sensors, a simple composition of still-images are taken at a predetermined frame rate, to create a video. This approach has several shortcomings, one such problem is data redundancy - each new frame contains information about all pixels, regardless of whether this information is new or not. Handling this unnecessary information wastes memory, power, and bandwidth \cite{Event-Based-Usefulness}. This problem might not be critical for the low frames per second (30-60 fps) use case, which is common when the video is intended for human observers. However, applications in computer vision, specifically ones that require real-time processing (more so even for high frames per second video), may suffer significantly from such inefficiencies \cite{Wang_2019_CVPR}.
        
         Several different sensors were suggested to overcome the shortcomings of the frame-based approach, some are bio-inspired (since it is known they can outperform traditional designs \cite{BioVision1,BioVision2}). Approaches such as optical flow sensors \cite{OpticalFlowSensor} (which instead of producing an image produces a vector for each pixel representing the apparent movement that was captured by that pixel), temporal contrast vision sensors \cite{TemporalContrastVisionSensor} (which only capture changes in pixel intensity, but use different hardware compared to DVS), and more \cite{ATIS, GSSVS, PoschQVGA} were suggested. However, the overwhelming majority of the market is still CCD and CMOS \cite{CCD-CMOS-Market}.
        
        Dynamic vision sensors (DVS) \cite{DVS-Dynamic-Range} are event-based cameras that provide high dynamic range (DR) and reduce the data rate and response times \cite{Event-Based-Usefulness}. Therefore, such sensors gain popularity these days \cite{DVS-response-model} and recently, several algorithms have been developed to support them \cite{dvs_applications1,dvs_applications2,dvs_applications3,dvs_applications4,dvs_applications5}.
        However, current technology is limited, especially in spatial resolution. Each DVS pixel works independently, and it measures log light intensity \cite{DVS-Log-Intensity}. Suppose it senses a significant enough change (more prominent than some threshold). In that case, it outputs an event that indicates the pixel's location and whether the event was an increase or decrease in intensity. The DVS sensor only measures the light intensity and therefore does not allow direct color detection. Hence, its vision is binary-like, only describing the polarity of the change in intensity. Due to its compressed data representation, the DVS is very efficient in memory and power consumption \cite{Amir_2017_CVPR}. Its advantage is even more prominent when comparing to high frames-per-second frame-based cameras. The reduced bandwidth required for recording only changes, rather than whole images, enables recording continuously at high temporal resolution, without being limited to very short videos. 
        
\begin{figure*}[h!]
            \centering
            \includegraphics[width=\linewidth]{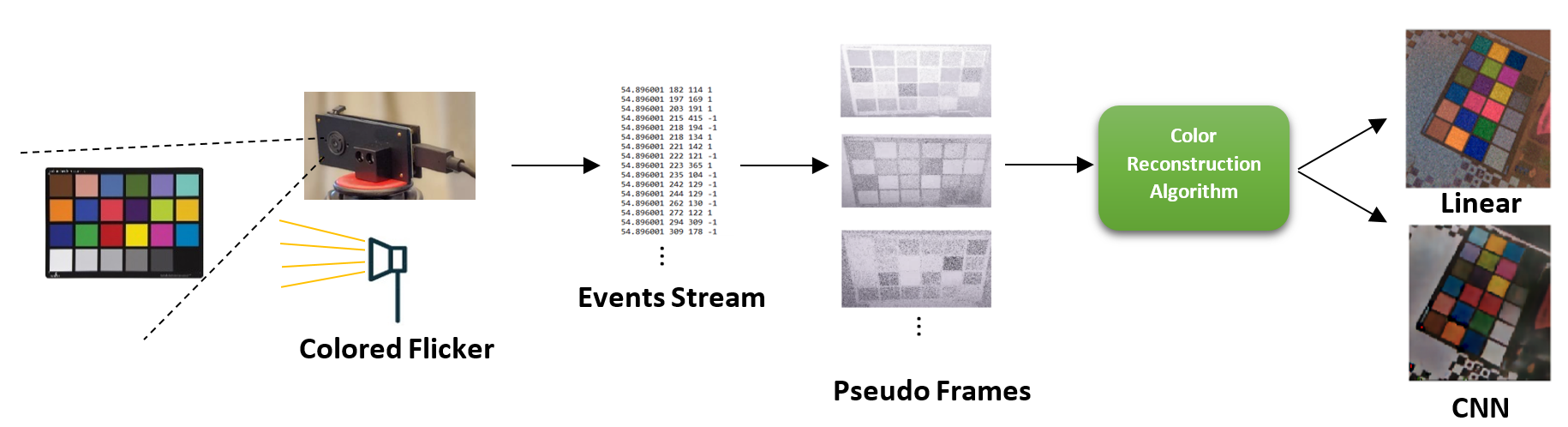}
            \caption{Reconstruction workflow. Left to right: The data is captured with a DVS sensor and a colored light source. Then, an event stream is created from the DVS, which is converted to Pseudo-frames representations. Finally, two different color reconstruction approaches can be applied.}
            \label{fig:reconstruction_workflow}
        \end{figure*}

        Color detection allows more information to be extracted from the DVS sensor. We focus on an algorithmic approach to reconstruct color from the binary-like output of the DVS sensor. Unlike current approaches \cite{color_event_camera}, ours does not reduce the native spatial resolution of the sensor. Our algorithm is based on the responses of the DVS sensor to different impulses of an RGB flicker. A list of frames is constructed from the events generated as a response to the flicker, from which features are extracted to reduce redundancy. Two feature extraction methods are described here, and each fits a different reconstruction method. The features are either used as input for a convolutional neural network or a linear estimator, depending on the discussed reconstruction method. The output of these algorithms is a single RGB frame in standard 8 bit color depth and the same spatial resolution as the DVS frames used as input, this output is a reconstructed frame of the scene in color. It could then be compared to a frame produced by a traditional camera. Fig. \ref{fig:reconstruction_workflow} describes the workflow presented in this paper. Our contributions:
        \begin{enumerate}
            \item A fast, real-time, linear method for reconstructing color from DVS camera.
            \item A CNN-based color reconstruction method for DVS.
            \item Non-linearity analysis of the DVS-flicker system and investigation of the DVS response to non continuous light sources.
        \end{enumerate}

    \section{Related Work}
        In traditional video cameras, color vision is achieved using a simple Color Filter Array (CFA) \cite{CFA-Is-Common} overlaid on the sensor pixels, with the obvious downside of reducing the spatial resolution of the sensor. In this approach, the output is divided into different channels, one for each filter color. For instance, in the case of the popular Bayer filter, it generally has one red and one blue channel and two green channels for a 2x2 binning configuration \cite{Bayer-Filter-Binning}. These channels can be used directly (with no interpolation) in frame-based cameras to produce a colored image or video. One might expect this approach to work just as well for event-based cameras. However, results show that naively composing the different color channels into an RGB image produces results that suffer from high salt-and-pepper type noise and poor color quality (see figure 4 subfigure C in \cite{CTCSDVS} for example). A more sophisticated approach for color interpolation from the different color channels, such as the ones employed in \cite{CTCSDVS, manifold, filter} (classical algorithms or filters) \cite{NNreconstruction} (neural network solution) produce better results, especially in terms of noise, but still suffer from poor color quality (see comparison between \cite{CTCSDVS} and our method in figure \ref{fig:comparison}).
        
        In previous works, the DVS data is assumed to have been produced from a continuous change in lighting. In contrast, in this work, we focus on color reconstruction using the approximately impulse response of a DVS-flicker system. Furthermore, the DVS exhibits interesting behavior under non-continuous light changes that we will discuss in this paper, which was not reported in the literature.
    
    \section{Dynamic Vision Sensor (DVS) and Setup}
        DVS cameras asynchronously output the position of a pixel that experienced a change in logarithmic intensity greater than some threshold \cite{DVS_PATENT}. This method of recording video has several advantages over more traditional synchronous sensors with absolute (as opposed to logarithmic) intensity sensitivity. For example, DVS cameras enjoyed a higher DR and a compressed data acquisition method, allowing for a more extraordinary ability to detect movement in poorly controlled lighting while using less power, less bandwidth, and much better latency.
        \subsection{DVS Operation Method}
            Pixels in a DVS sensor contain photoreceptors that translate incident photons to current. Transistors and capacitors are then used to create a differential mechanism, which is triggered only when the incident log luminosity difference is greater than a threshold \cite{DVS-Structure}.

        \subsection{DVS Response Analysis}
            One can model the response of the DVS sensor to a change in brightness as a delta function:
            \begin{equation}
                \delta (\textbf{r} - \textbf{r}_i ,t-t_0)
            \end{equation}
  
            Where $ \mathbf{r_i} $ is the pixel at which the event has occurred, and $t_0$ is the time at which the event has occurred.
            The sensor responds to changes in the log intensity of each pixel, which can be modeled for some pixel at time $t_k$ as\cite{DVS-response-model}:
            \begin{equation}
                \Delta L ( \mathbf{r_i},t_k ) \geq p_k C
            \end{equation}
            Where $L \equiv log(I)$, and
            \begin{equation}
                \Delta L(\mathbf{r_i},t_k) \equiv L(\mathbf{r_i},t_k)-L(\mathbf{r_i},t_k-\Delta t_k)
            \end{equation}  
            Here $I(\mathbf{r_i},t_k)$ is the intensity, $p_k$ is the polarity of the pixel, which is $+1$ for an increase or $-1$ for a decrease in brightness of that pixel. 
            The variable C corresponds to the threshold that allows for a response to be observed and is derived from the pixel bias currents in the sensor.
            
      
        \subsection{Creating Video From Bitstream}
            DVS outputs a bitstream using address-event representation (AER), each event detected by the DVS has an address (which represents the position of the pixel that detected the change), polarity, one if the change detected was that the brightness increased,  -1 if the change was the brightness decreased and the time the event was detected. In order to turn this list of events into a video, we first choose the desired FPS (we chose 600 for best performance with our specific DVS model, but it is possible to work with up to 1000 fps, and newer models can go even higher), after that choice we uniformly quantize time, and to create a frame we sum all the events that occurred in each time slice to a single frame, which holds the data about the total event count in each pixel at a time period reciprocal to the fps. A similar temporal binning procedure has been introduced in \cite{CTCSDVS}.

    \subsection{System Setup}
            We used Samsung DVS Gen3 \cite{dvs_gen3}, and positioned it such that it was facing a static scene 14 inches (35.5 cm) away. For the flicker we used a screen capable of producing light of different wavelengths and intensities. We placed the flicker directly behind the camera facing the scene, such that it is illuminating it relatively approximately uniformly, because it is bigger in area then the region of the scene captured by the DVS. The flicker changes the emitted color at a 3 Hz frequency. For the calibration process (will be discussed) we used PointGrey Grasshopper3 U3, RGB camera adjacent to the DVS.
            
            The scene is static, so the camera can not see anything if the flicker does not change color. The flicker's light is reflected off the scene and into the sensor, meaning that if we look at a single DVS pixel it measures whether the temporal change in the integral over all frequencies of the product of the spectrum, the reflection spectrum of the material and the quantum efficiency of the sensor is greater than some threshold. When light from the source changes in color or intensity, that change is recorded in the DVS.
            
            \begin{figure}[h!]
                \centering
                \includegraphics[width=1\linewidth]{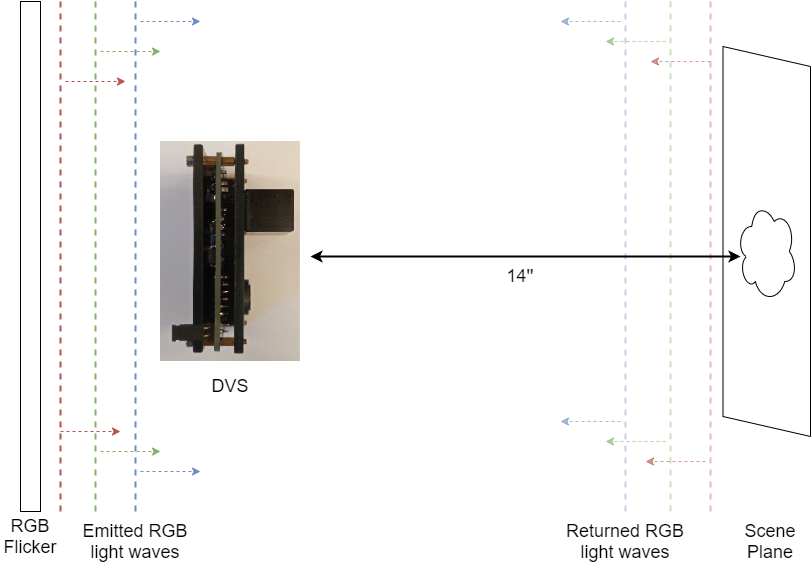}
                \caption{System schematic. The distance from the DVS to the flicker is much shorter than 14''. The RGB flicker emits light at a 3 Hz frequency}
                \label{fig:system_setup}
            \end{figure}    
            
            Using this system we intend to capture a bitstream generated by the DVS as a response to the flicker, from which a single RGB frame is produced. This frame is an RGB image of the scene that is the same resolution as the original DVS video. We present here two algorithmic approaches (one linear and one using CNN, in next section) for producing that RGB frame.
            We will create a method of feature extraction for each of the two different algorithmic approaches, and in order to train the CNN, we will create a labeled data-set using a stereoscopic system with DVS and a standard frame-based camera. Finally we will provide an explanation about our methods performances and shortcomings.
    
    \section{Method - Linear Approach}
        We introduce here a fast, real-time linear method for creating an RGB frame. This method will estimate the color of each pixel based only on the single corresponding DVS pixel. Therefore our problem is simplified to reconstructing a the color of a single pixel from a list of DVS events for that pixel only, we will use the same method for all pixels to create a full RGB image. As will be explained in the following section, we further reduce the problem to estimating the color from a vector of real positive numbers. We generate a few labeled vectors and then build the Linear-Minimum-Mean-Square-Error (LMMSE) estimator using the Moore-Penrose pseudo-inverse \cite{bishop}.
        
        \subsection{Feature Extraction}
            When recording the scene for color reconstruction, the intensity and the color of the flicker are varied. Therefore, after changing the intensity of light to a new one, the new intensity is treated as if it is a different color (regarding feature extraction).
            
            Pre-processing the DVS output yields a list of frames, Each being the sum of events that occurred in a particular time slice. We start by choosing the time intervals, each corresponding to the response period of the DVS, to a change in the reflected intensity outside the scene, which will be later referred to as integration windows. Thus, a response curve of each pixel to each change in color is yielded. We use a flicker that transmits three different colors (RGB) at three different intensities. 
            
            In order to reduce the size of the data, we use the average of the response to each color change (a single number per pixel) over a pre-defined integration window, and responses corresponding to the exact color change are averaged. The result is a vector of length $N+1$ (where $N=9$ is the number of color changes the light source provides, and in addition, there is a bias parameter) for each pixel in the 640 by 480 sensor array. In order to justify this, we will approximate each DVS pixel as an independent LTI system.
            
            \subsubsection{LTI Approximation}
                When the flicker is on, each pixel measures the light that is reflected from some part of the scene, suppose for a given pixel, that part of the scene is all the same color. When the flicker is on one color, the pixel will measure incident photon flux of $F_i$, when the flicker changes color (suppose at $t=0$), the incident photon flux changes to $F_f$. Thus, the DVS pixel will output SEVERAL events corresponding to this change, this is a unique feature resulting from the non-continuous change in light intensity. The amount of events is dependant on the size of change $\Delta F = F_f - F_i$. This property is crucial for distinguishing between colors of different brightness, such as different shades of gray, since the brightness of some colors is directly linked to the intensity of the reflected photons off of it. The stream of events originating from the flicker change lasts well after the flicker transition is over, meaning that if light intensity changes quickly, the DVS will not treat this intensity change as a single event, but rather will continue outputting events for some time, such that the number of events is proportional in some way to the change in intensity. 
                \begin{figure}[h!]
                    \centering
                    \includegraphics[width=1\linewidth]{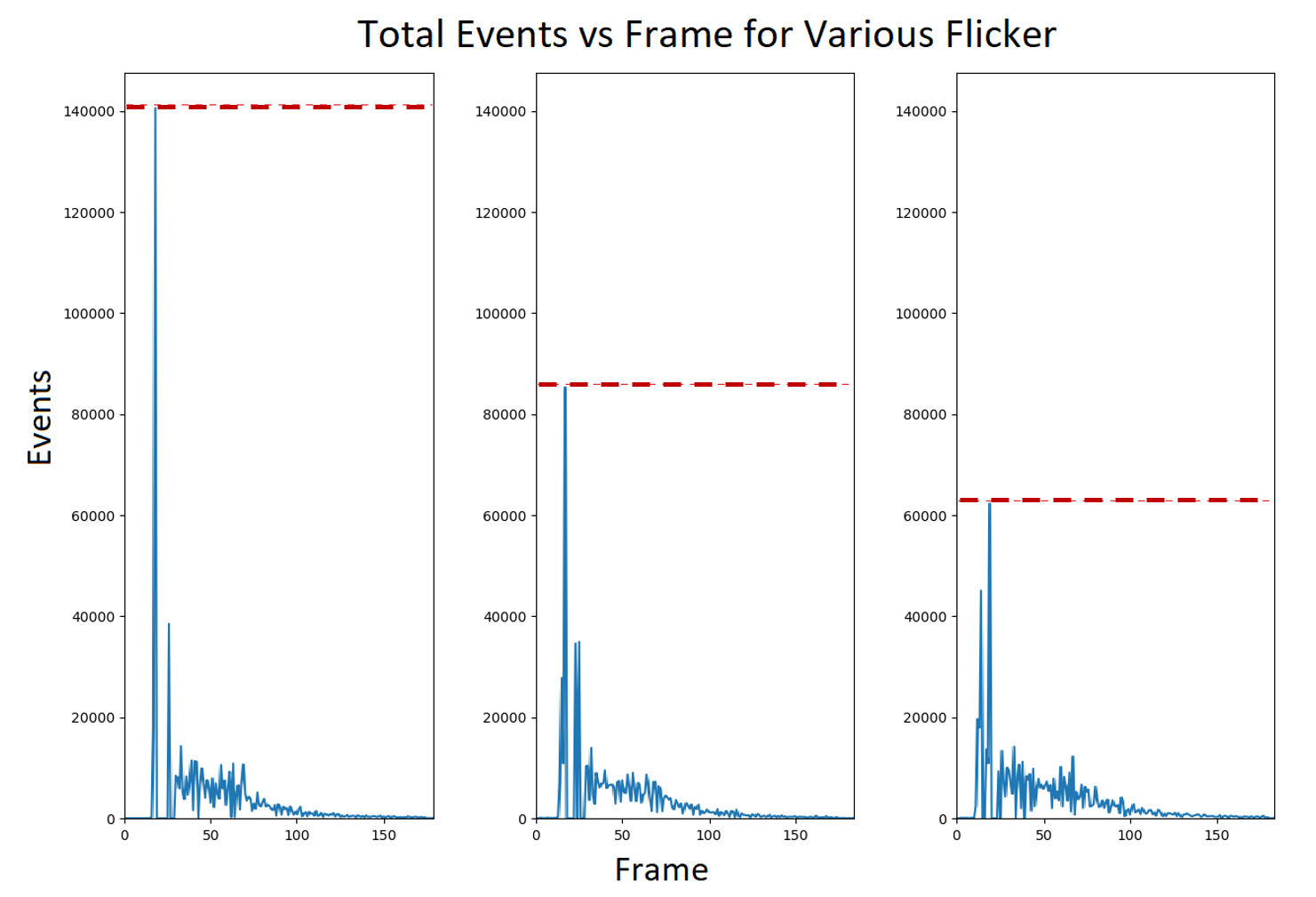}
                    \caption{DVS event count for gray-scale flicker with decreasing gray-scale intensities, left to right. The red dashed lines indicate the peak event count for each gray-scale flicker intensity, showing a correlation between the intensity of the flicker and the amount of events recorded. }
                    \label{fig:color_matrix1}
                \end{figure}

                Assuming the logarithm of the transmitted intensity from the flicker could be described as a step function $$f(t)=au(t)+u_0$$
                The output current of a DVS pixel is approximated as:
                $$ I_{DVS}(t) \propto ae^{-bt} $$
                The exponential form of the output current is implied from the discharge currents of the capacitors in each DVS pixel. We assume the probability of a DVS pixel to register an event is proportional to the output current:
                $$f_{DVS}(t) \propto I_{DVS}(t)$$
                As long as $I_{DVS}(t)$ is above some threshold, which allows the DVS sensor to record events only when the log intensity change is sufficiently significant. 
                
                The LTI approximation suggests a sufficient description of the DVS response to the flicker change is taking the number of events that occurred during the flicker change, i.e., integrating over time. Therefore, we suggest the following criterion for characterizing the DVS response to the flicker change:
                \begin{equation}
                    \label{LTI_integration_approximation}
                    F_{\lambda_1 \rightarrow \lambda_2} \approx \int\limits_0^\infty f_{DVS}(t)dt \approx \int\limits_0^\tau f_{DVS}(t)dt
                \end{equation}
                
                We clip the integration at a finite time $\tau$, chosen empirically, in order to approximate the integral. It is approximated as the time difference between the moment the flicker change has happened and the moment where $f_{DVS}(t)$ has decayed enough. In this work, these moments are identified by measuring the DVS pixels' event count over an entire frame with respect to time. A flicker change triggers a local maximum in the event count over time, which then decays proportionally to $f_{DVS}(t)$.
                An integration window $\tau$ is taken from the frame where the event count is at a local maximum to the frame at which the total event count equals the average event count over all frames. Noise and scattering are taken into account when determining the integration windows.

                A pixel of a particular color reflects different amounts of the incident photons transmitted by the flicker, depending on their respective wavelengths. Thus, changing the flicker's color causes the object the pixel represents to reflect a different amount of light and the observed intensity changes. This causes a reading by the DVS sensor. Based on these readings, we suggest it is possible to reconstruct the RGB color profile of the observed scene. 
                
                The LTI integration approximation justifies using the LHS of equation \ref{LTI_integration_approximation} as the components of the feature vector, but if we take noise into account, this approximation begins to break down, and the use of spatial correlation is needed for improved approximation.
                \begin{figure}[!h]
                    \centering
                    \includegraphics[scale=0.185]{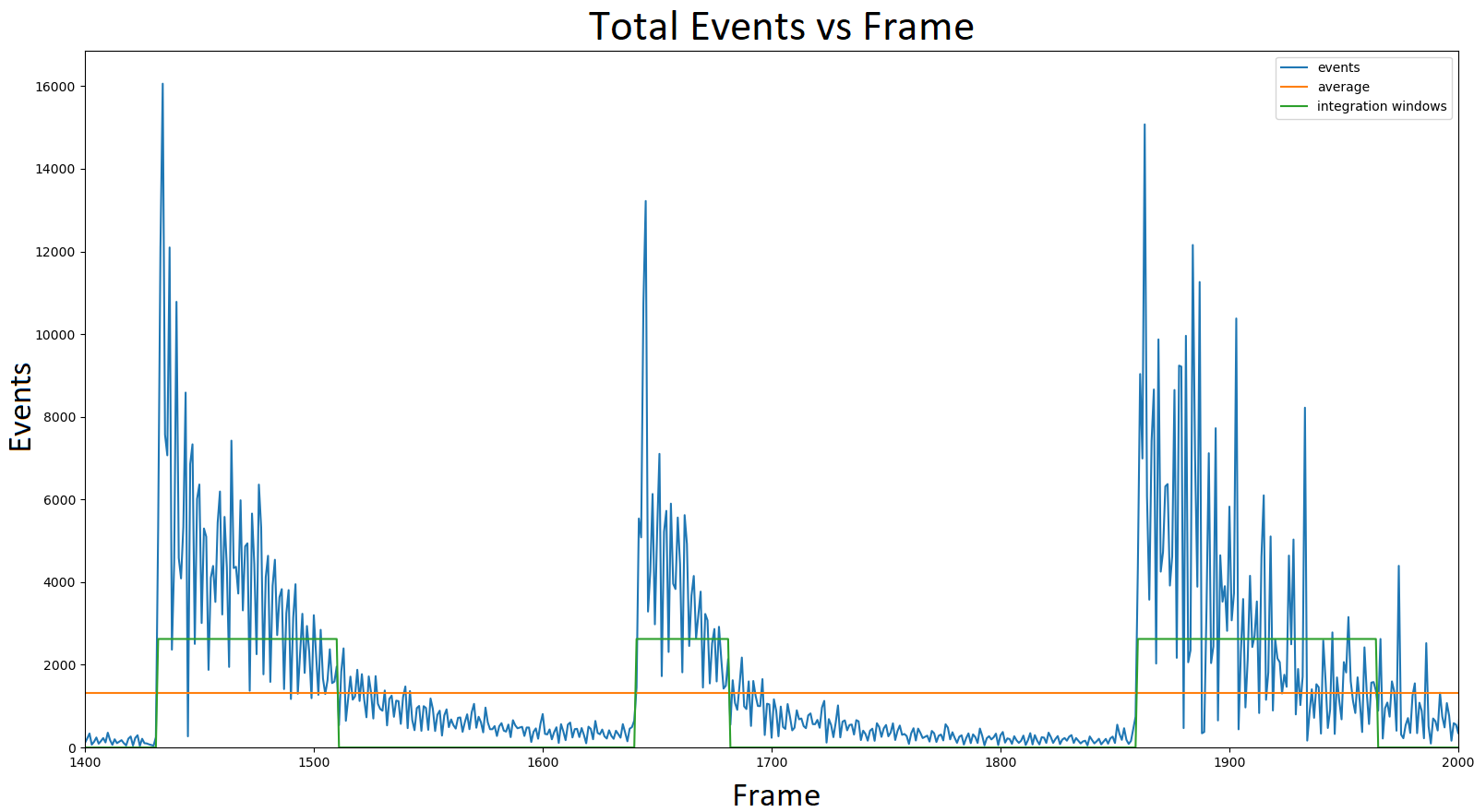}
                    \caption{Amount of DVS events in a frame vs frame number. The measurements were recorded while RGB flicker changes were taking place. The integration windows depict the frames with enough events caused by a flicker color change.}
                    \label{fig:total_events_vs_frame}
                \end{figure}
                
                The exponential form of the DVS event count can be seen in fig. \ref{fig:total_events_vs_frame}, which supports our model of the DVS pixel response probability $f_{DVS}(t)$. The noisy response shown in fig. \ref{fig:total_events_vs_frame} is a result of light scattering in addition to the inherent noise in the sensor.

    \section{Method - CNN Approach}
        The linear approach suffers from low SNR and low color fidelity.
        The shortcomings of the linear estimator are due to several factors but mainly the sensor noise (thermal noise is significant for this sensor because it has a large surface area compared to modern CMOS or DSLR sensors) and inherent nonlinearities of the device, which were ignored in the last chapter's analysis. The sensor in this work has a thermal noise of 0.03 events/pix/sec at 25\degree C \cite{DVS-response-model}. An additional form of noise is shot noise, which is significant under low ambient light conditions, such as in our setup. The low color fidelity of the linear approach suggests the LTI assumption falls short in yielding a highly accurate color reconstruction. Thus, a different, more robust method should be employed for this problem. A natural solution is a convolutional neural network, as it is common in image processing settings. This method uses a non-linear estimator for the color, and in addition, it takes into account the inherent spatial correlation in the DVS output to reduce output noise. 
        
        The input to the network is 288 frames (32 frames for each of the nine flicker transitions) from the DVS, selected to contain the most information about the flicker transition. Since the network is fully convolutional, different spatial resolutions can be used for the input, but the output must be of appropriate size. 
        
        From looking at the output of the DVS, a few things are clear. First, it is sparse; second, it is noisy; and third, there is a lot of spatial and temporal correlation between pixels. This matches previous works' findings of the DVS event stream \cite{DVS-response-model}. 
        In addition, the linear approximation has problems distinguishing different shades of gray (among other problems), using the spatial and temporal correlation of the data will help produce better results, as well as help deal with a certain design flaw of the sensor, it seems that certain pixels respond because neighboring pixels responded and not because they sensed a change in photon flux, this is non-linear behavior that cannot be accounted for using a simple linear approximation (see figure \ref{fig:dvs_wave} in appendix).
        \subsection{CNN Architecture}
            The network architecture is fully convolutional, inspired by U-Net \cite{UNET} and Xception \cite{XNET}. Similar to U-Net, this network consists of a contracting path and an expanding path. However, it also includes several layers connecting the contracting and expanding part used to add more weights that improve the model. 
            
            Each layer in the contracting path reduces the spatial dimensions and increases the number of channels using repeated Xception layers \cite{XNET} based on separable convolutions. Each layer in the expanding path increases the spatial dimensions and reduces the channels using separable transposed convolutions. In the end, we get back to the desired output size (in our case, it will be the same as the input size), and the channels will be reduced down to 3 channels, 1 for each RGB color. The path connecting the contracting and expanding layers does not change the size of the data.
        
        \subsection{Loss Function}
            The loss function is a weighted average of MS-SSIM \cite{MS-SSIM} and $L_1$ norm.
            \begin{equation} \label{eq:loss}
                \mathcal{L}(Y,\hat{Y}) = 0.8 ||\hat{Y}-Y||_1 + 0.2  L_{\text{MS-SSIM}}(\hat{Y},Y)
            \end{equation}
            While $\hat{Y}$ and $Y$ represents the reconstructed and real image, respectively.
            The coefficients of the different components of the loss function were tuned using hyperparameter optimization. Other losses were tested, including L2 loss and L1 loss that compares only the hue and saturation (without the lightness) of the images, or only the lightness without the hue and saturation, but none outperformed the loss we chose. 
            
            The SSIM index is a distortion measure that is supposed to more faithfully represent how humans perceive distortions in images, assuming that the way human visual perception works depends significantly on extracting structural information from an image. It is helpful for us because the linear approach (despite being noisy and not producing the most accurate colors) did seem to be able to produce images that have the same structure as the original one.

        \subsection{Training}
            Data labels are acquired using a dual-sensor apparatus composed of DVS and RGB sensors. For the RGB sensor we used a PointGrey Grasshopper3 U3, camera, with resolution of 1920 × 1200 (2.3 MP), fps of 163 and 8 bit color depth for each of its 3 color channels. A calibration process yields matching sets of points in each of the sensors using the Harris Corner Detector algorithm, which is then used to calculate a holography that transforms the perspective of the RGB sensor to the perspective of the DVS sensor.
            
            The calibration process assumes the captured scene is located in a dark room on a plane at a distance of 14'' from the DVS sensor. Therefore, training data are taken on 2D scenes for preserving the calibration accuracy.
            Each training sample contains a series of frames, most of which hold the responses of the scene to the changes in the flicker, and a minority of the frames are background noise frames before the changes in the flicker. For example, in the case of an RGB flicker with three intensities, we use 32 frames per color and intensity, totaling 288 frames.

    \section{Results}
        \subsection{Linear Approach}
\begin{figure}[h!]
                \centering
                \includegraphics[width=0.5\linewidth]{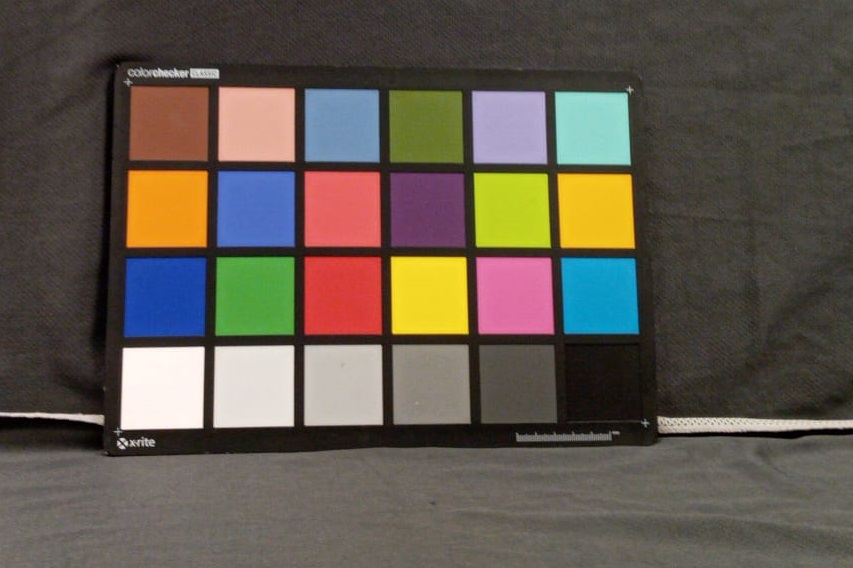}
                \includegraphics[width=0.5\linewidth]{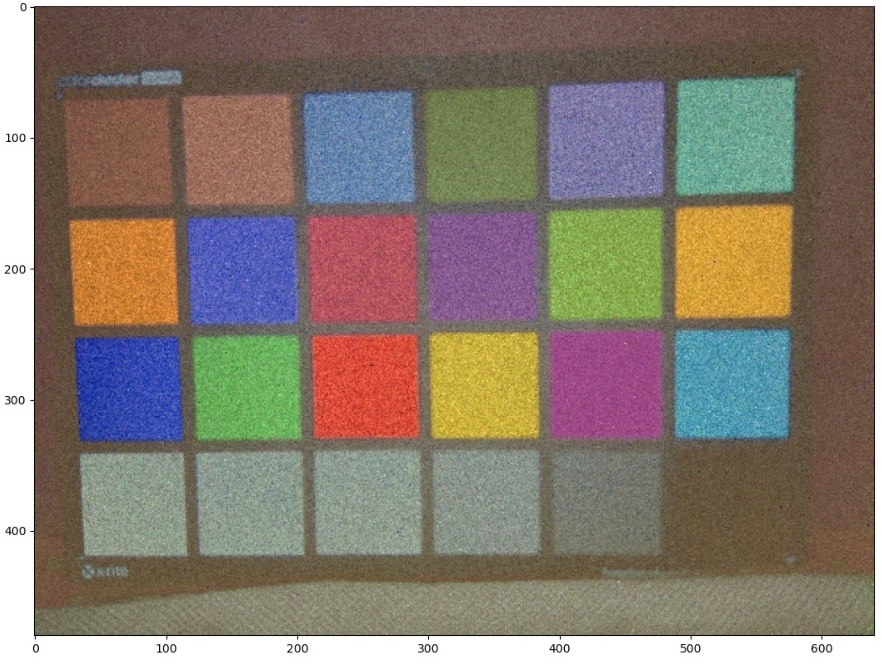}
                \caption{Top:X-Rite color matrix. Bottom: color reconstruction using the 9 flicker (3 colors, 3 intensities) linear approach.}
                \label{fig:color_matrix_reconstruction}
            \end{figure}

            Some linear reconstructions are shown in figure \ref{fig:color_matrix_reconstruction}:
            The linear approach is shown to reconstruct color, although it is very noisy. This is the result of the pixel-wise approach implemented here, where the spatial correlation between the colors of neighboring pixels is ignored. The fact that noise causes neighboring pixels to experience different event readout patterns causes neighboring pixels to have different feature vectors. Therefore their reconstructed colors are not similar. This can be rectified by considering the spatial correlation between neighboring pixels, as the CNN approach does.
            
\begin{figure}[h!]
                \centering
                \includegraphics[width=0.5\linewidth]{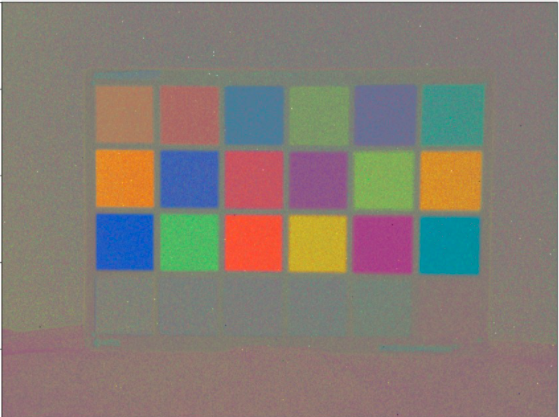}
                \caption{X-Rite matrix reconstruction with a single intensity flicker. The gray colors are almost indistinguishable and the color fidelity has deteriorated, compared to the three-intensity linear reconstruction.}
                \label{fig:color_matrix_reconstruction_single_intensity}
            \end{figure}

            We have also checked that transmitting light in multiple intensities is crucial for distinguishing between different shades of colors by only recording the DVS responses to three changes in the flicker.
            The result is shown in figure \ref{fig:color_matrix_reconstruction_single_intensity}.
            Figure \ref{fig:color_matrix_reconstruction} shows that using three intensities enables a differentiation between the gray colors in the X-Rite color matrix. However, as seen in fig. \ref{fig:color_matrix_reconstruction_single_intensity}, using a flicker that transmits RGB light in a single intensity makes the gray colors almost indistinguishable, and is detrimental to the accuracy of the reconstruction. All gray colors have RGB values proportional to $(1,1,1)$, i.e. a gray color has the same value in each RGB channel, indicating the proportion coefficient, or the intensity. Since the DVS output is binary in essence, different gray colors on the scene will respond similarly to the color changes in a flicker with a single intensity.
            Training with such a setup will result in intensity-mismatched reconstructions. As seen in fig. \ref{fig:color_matrix1}, the amount of events recorded depends on the intensity of the flicker relative to the intensity of the scene. Therefore, one can quantify the relative intensity of each pixel by recording its responses to different flicker intensities. Thus,  Multiple intensities allows to mitigate the aforementioned problem by measure the relative intensity of each actual pixel on the scene to the projected intensity of the flicker.

        \subsection{CNN Approach}
\begin{figure}[h!]
                \centering
                \includegraphics[width=1\linewidth]{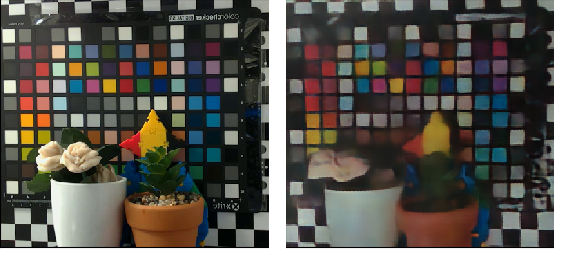}
                \includegraphics[width=1\linewidth]{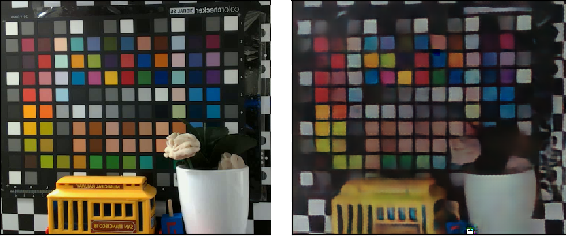}
                \caption{Our model's reconstruction of 3D scene. Left: The original image, right: Our CNN-based model reconstruction. The RMSE for the top reconstruction is 45 and for the bottom is 47}
                \label{fig:3d_reconstruction}
            \end{figure}

            Figure \ref{fig:3d_reconstruction} shows some of our color reconstruction results on 3D scenes that are 14'' away from the DVS sensor (distance is measured from the DVS lens to the center of the 3D scene). The resulting reconstruction shows high fidelity and low noise. Thus, this solution is viable for colors reconstruction of 3D and 2D scenes. Small details might not be resolved by this approach due to the smoothing effect of the CNN. 
            
            \subsubsection{Robustness}
            Training data have been captured under fixed lab conditions - in particular, the captured scene is at a distance of 14'' from the DVS sensor, and the scene is situated in a dark room. Therefore, the robustness of the system to changes in these conditions is shown.
            
\begin{figure}[h!]
                \centering
                \includegraphics[width=1\linewidth]{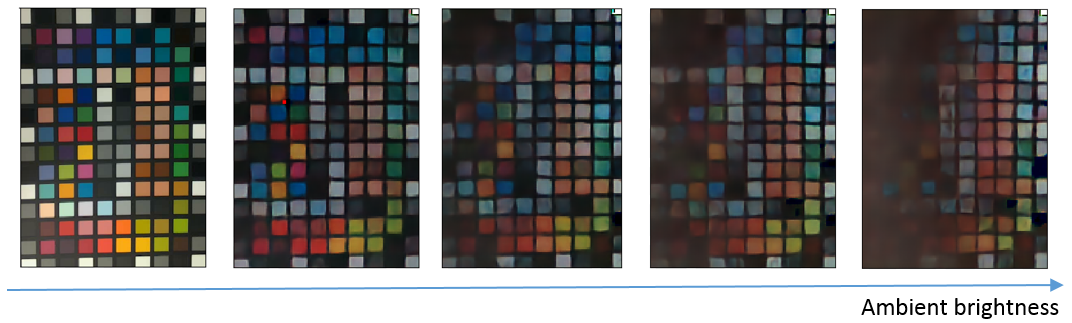}
                \begin{center}
                    \begin{tabular}{ |c|c| } 
                        \hline
                        Brightness [lux] & Normalized Loss \\ 
                        \hline
                        0.4 & 1  \\ 
                        \hline
                        10.8 & 1.13  \\ 
                        \hline
                        21.2 & 1.18 \\
                        \hline
                        39.2 & 1.29 \\
                        \hline
                    \end{tabular}
                \end{center}
                \caption{Top: Reconstruction for different ambient light conditions. RGB ground truth is shown in the leftmost picture. Bottom: relative loss for each reconstruction. Loss is calculated as in equation \ref{eq:loss}.}
                \label{fig:robustness_brightness}
            \end{figure}

            The robustness to changes in the ambient brightness has been studied by using a light source behind the flicker, directed at the scene, which is an extended color matrix. The flicker obstructs some of the incident light from the light source, and therefore regional intensity variance in the scene has been observed. The light intensity at the scene plane is taken as the average intensity measured at the corners of the color matrix. As seen in figure \ref{fig:robustness_brightness}, brighter ambient light causes worse color reconstruction. The saturation of the scene by enough ambient light causes the flickering color and intensity changes to have a less significant change in the incident intensity measured by the DVS pixels, and therefore fewer events are recorded. 

\begin{figure}[h!]
                \centering
                \includegraphics[width=1\linewidth]{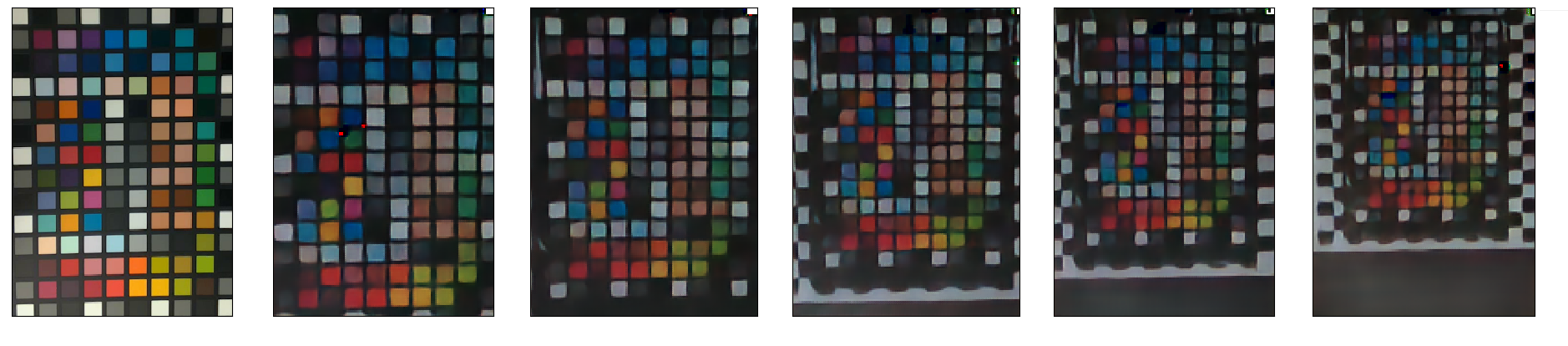}
                \caption{Our CNN-based model reconstruction results for different distances. Distances from left to right: $5.08_{cm}$, $11.176_{cm}$, $17.272_{cm}$, $23.368_{cm}$, $29.464_{cm}$ and $35.56_{cm}$.}
                \label{fig:robustness_distance}
            \end{figure}

            Distance robustness is also studied, where the color matrix was situated in a dark room, with the same ambient brightness as in the training sessions. Measurements were taken starting at a distance 14'' from the DVS and increasing by 2'' increments. As seen in figure \ref{fig:robustness_distance}, the reconstruction capabilities of the CNN approach are not limited to the fixed training distance but allow for a distance generalization.
            
            \subsubsection{Ablation Study}
\begin{figure}[h!]
                \centering
                \includegraphics[width=1\linewidth]{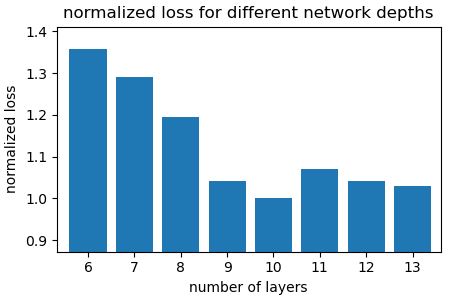}
                \caption{Our Neural-Network normalized loss for different number of layers. We refer this analysis (without a formal proof) an interpretation for the DVS non-linearity degree.}
                \label{fig:loss_vs_depth}
            \end{figure}

            It is worth investigating the optimal depth of the network. In order to accomplish that, we trained several variations of the network with different depths on the same data (and the same number of epochs), and we got the following results shown in figure \ref{fig:loss_vs_depth}
            
\begin{figure}[h!]
                \centering
                \includegraphics[width=1\linewidth]{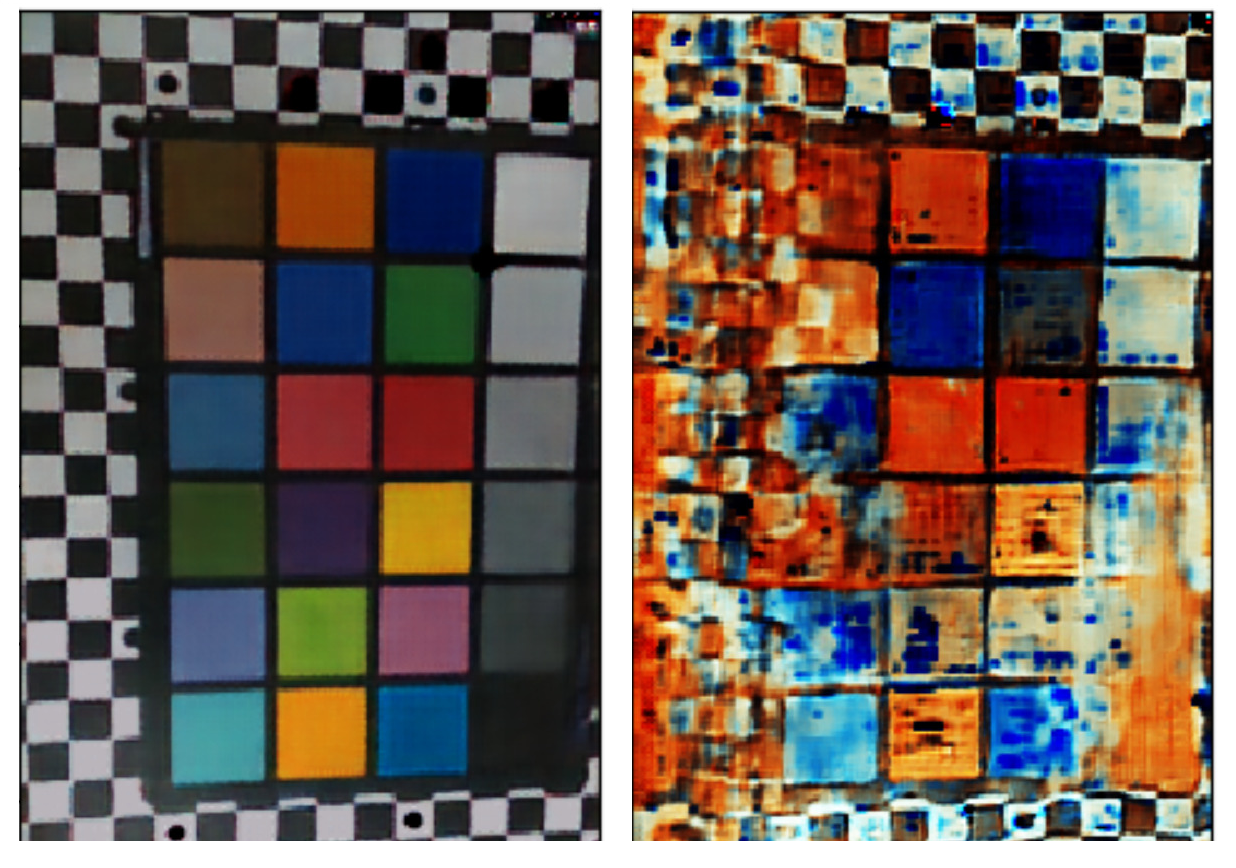}
                \caption{Using the DVS response for different flicker intensity improve significantly the reconstruction quality. Left: Our CNN-based model reconstruction using 3 different intensities and 3 different colors. Right: Same model's reconstruction using a single intensity and 3 colors.}
                \label{fig:intensities_compare}
            \end{figure}

\begin{figure}[h!]
                \centering
                \includegraphics[width=1\linewidth]{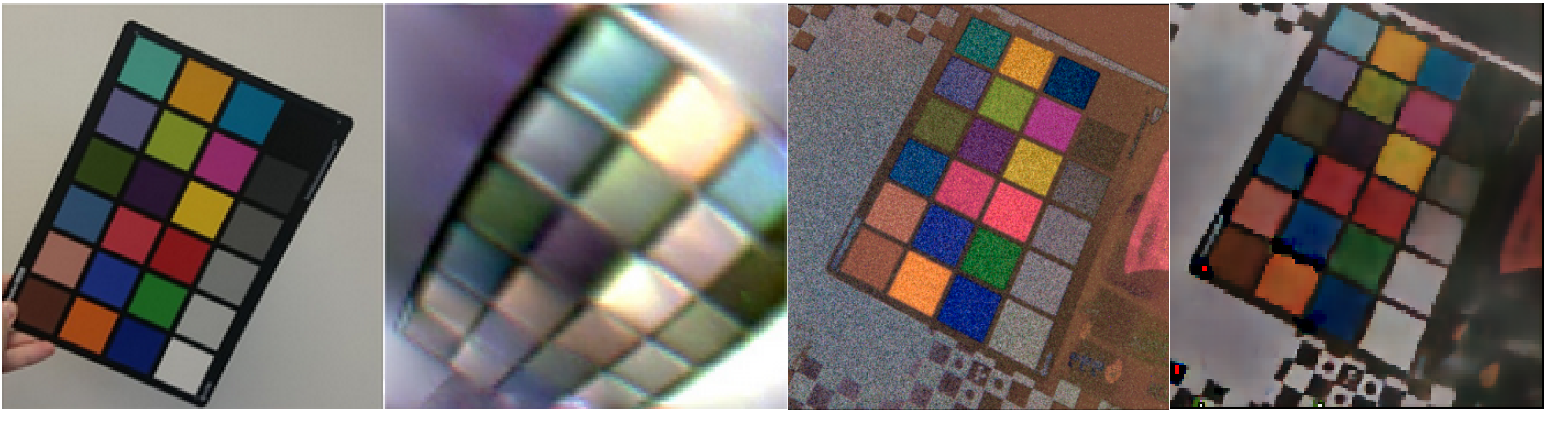}
                \begin{center}
                    \begin{tabular}{ |c|c|c| } 
                        \hline
                        Perceived Color & Our MSE & MSE of \cite{CTCSDVS} \\ 
                        \hline
                        (4,138,122) & \textbf{6368} & 17476  \\ 
                        \hline
                        (111,129,39) & \textbf{749} & 8157  \\ 
                        \hline
                        (180,135,18) & \textbf{662} & 43097 \\
                        \hline
                        (165,32,42) & \textbf{566} & 13030 \\
                        \hline
                        (144,3,8) & \textbf{734} & 19043 \\
                        \hline
                        (68,69,64) & \textbf{237} & 9269 \\
                        \hline
                    \end{tabular}
                \end{center}
                \caption{Left to right: X-Rite color matrix, reconstruction by \cite{CTCSDVS}, our linear reconstruction, our CNN reconstruction}
                \label{fig:comparison}
            \end{figure}

            An additional component that is worthwhile to study is the use of different intensities for the flicker (we used 3 colors with 3 different intensities), what would happen to the reconstruction if we only used 1 intensity for the flicker. To that end we trained the same network architecture for 2 different data sets, the first set in the normal one where we used a 3 color flicker with 3 different light intensities for each color, and the second contains the same images, but this time a 3 color flicker was used with 1 intensity per color. In figure \ref{fig:intensities_compare} we can see that the reconstruction failed to produce the right colors.
            
            Comparing our results to those of \cite{CTCSDVS} one can see that our method is superior in terms of color reconstruction accuracy, at least on the X-Rite color matrix. The perceived colors of the Xrite color matrix and the respective MSE values are calculated using 4-point averages of the images from \cite{CTCSDVS} and our reconstruction of the same color matrix. In addition, in contrast to color filter array (CFA) based color reconstructions, our methods keep the spatial resolutions of the resulting reconstructions close to the native resolution of the DVS sensor.
            The method presented in \cite{CTCSDVS} uses DAVIS (Dynamic and Active Vision Sensor) pixels with CFA, which means they are using more sensors to create the same resolution pixel as in our method.

    \section{Discussion}
        
        The system described in this work does manage to create a colored image using the protocol that we presented. However, it has its limitations: reconstruction accuracy depends heavily on ambient light conditions - if different ambient light conditions are taken into account during training, the reconstruction will be less sensitive to ambient light conditions; the flicker's intensity and dynamic range affect the reconstruction quality. That being said, the system does have the ability to reconstruct color from images outside the training data, as can be seen in figure \ref{fig:3d_reconstruction}. One can overcome the latter limitation by changing the intensity levels of the flicker. In addition, our system does not allow color reconstruction of a non-stationary scene since the response time of the DVS pixels to a flicker color change is non-zero. Therefore, the time it takes for a flicker cycle to be performed limits the timescale of changes in the scene using our approach. The critical distinction is between events generated by flicker changes and events generated by movements in the scene. An optical flow algorithm combined with a CNN can be used to overcome this limitation.
        
        Using a DAVIS sensor would also most likely improve the performance and ease the calibration method (the active pixels could be used to create a regular still image used for calibration). In addition, since the DVS has no sense of absolute light intensity (only relative), using a DAVIS could improve performance for different ambient light conditions.

    \section{Conclusions}
        In this work, we showed a novel approach for DVS color reconstruction - using an RGB flicker to characterize the sensor's responses to different colors of the flicker, even though the sensor provides only a binary indication of intensity change (events). By emitting RGB light and changing the flicker's color in a non-continuous fashion, we can figure out the scene's spectral data. Two methods for reconstruction were described, each with a different success rate. While both methods succeed in the reconstruction, the CNN approach has higher fidelity and less graininess than the linear method. This is somewhat expected due to previously described non-linear effects in the system.
        Our solution can be utilized to sample the spectrum with more than three different flicker color changes by including different flicker outputs. This allows for creating a hyperspectral image of the scene. 
        This implementation outperforms a CFA-based reconstruction since the latter only samples wavelengths at a few specific wavelengths. At the same time, an RGB flicker allows for sampling wavelengths anywhere on the optical spectrum.
        
    \bibliographystyle{plain}
    \bibliography{references}
    
    \newpage
    \section{Appendix}
        \begin{figure}[h!]
                \centering \includegraphics[width=1.0\linewidth]{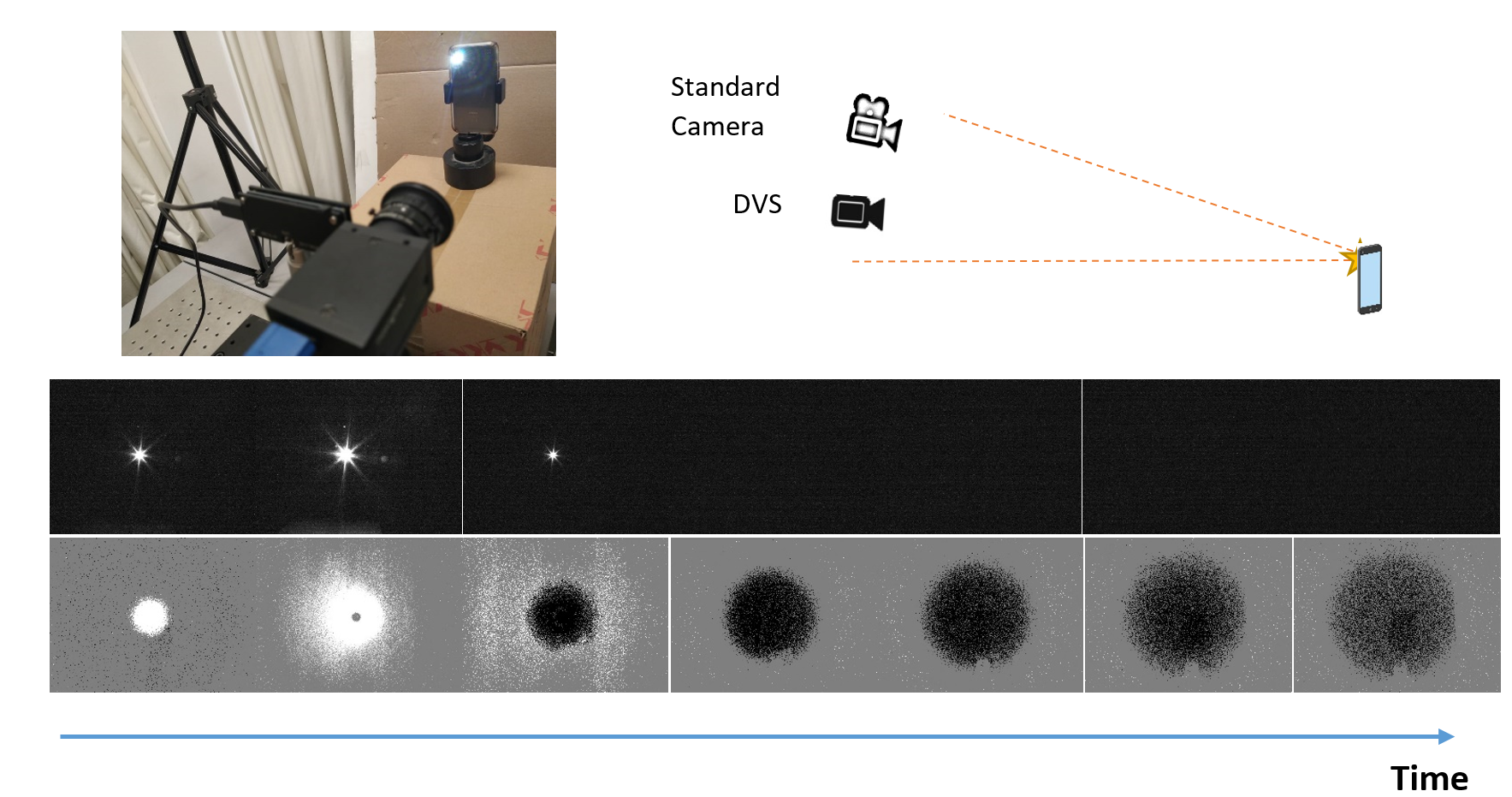}
                \caption{Example of non-linear behavior in the form of wave-like ripple of events that shouldn't occur}
                \label{fig:dvs_wave}
            \end{figure}
            
        \begin{figure}[h!]
            \centering
            \includegraphics[width=1.0\linewidth]{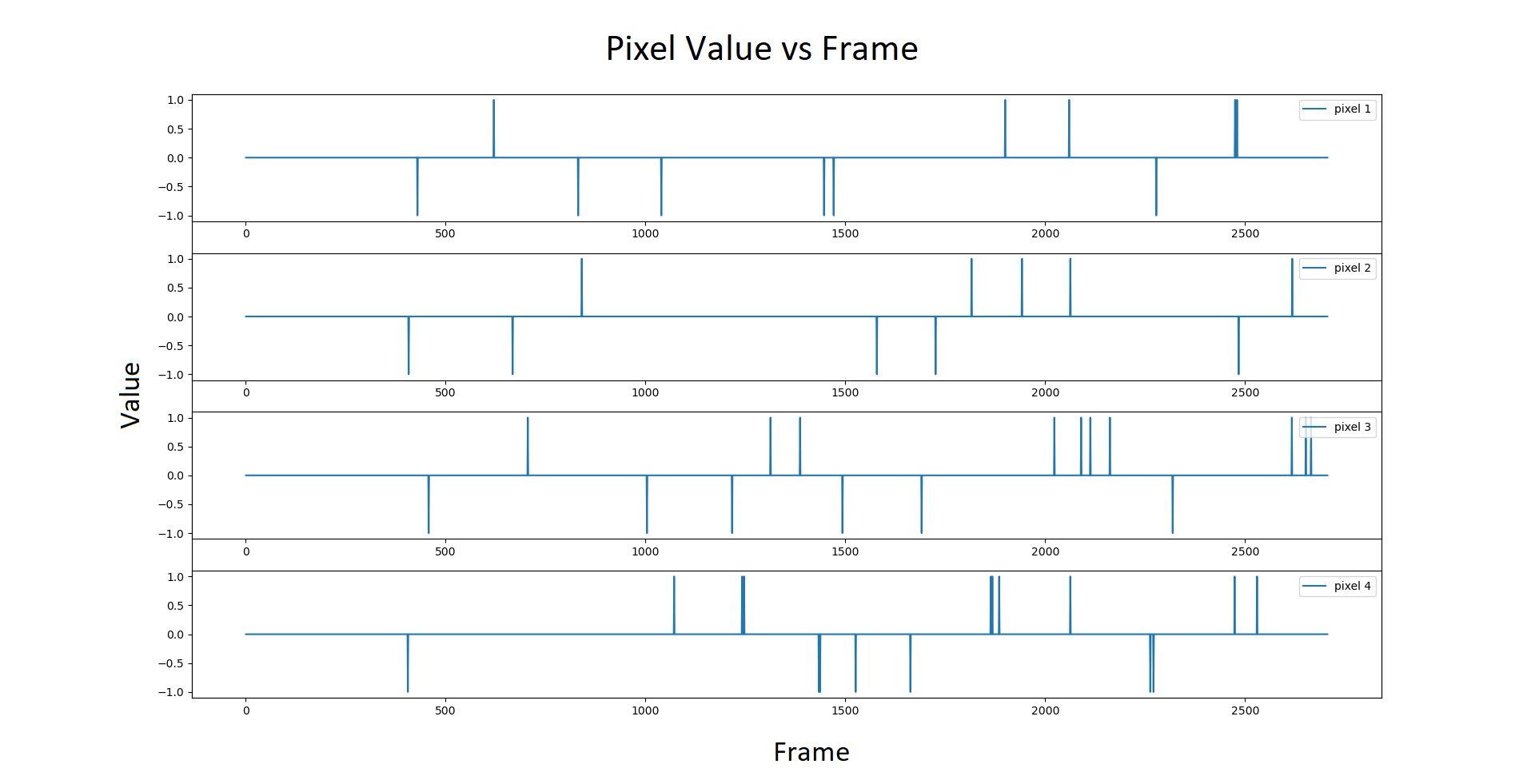}
            \caption{DVS pixel responses for different pixels, showing the response is sparse and that not all pixels respond simultaneously.}
            \label{fig:sparseness}
        \end{figure}

\end{document}